\documentclass[preprint,aps,tightenlines,floatfix,showpacs]{revtex4}
\usepackage{graphicx}
\usepackage{bm}
\usepackage{amsmath}
\usepackage{amssymb}

\begin{document}

\title{Compound and quasi-compound states in low-energy scattering
of nucleons from ${}^{12}$C.}

\author{G.~Pisent{$^{(1)}$}}
\email{gualtiero.pisent@pd.infn.it}
\author{J.~P.~Svenne$^{(2)}$}
\email{svenne@physics.umanitoba.ca}
\author{L.~Canton$^{(1)}$}
\email{luciano.canton@pd.infn.it}
\author{K.~Amos$^{(3)}$}
\email{amos@physics.unimelb.edu.au}
\author{S.~Karataglidis$^{(3)}$}
\email{kara@physics.unimelb.edu.au}
\author{D.~van~der~Knijff$^{(4)}$}
\email{dirk@unimelb.edu.au}
\affiliation{
$^{(1)}$Dipartimento di Fisica dell'Universit\`a
di Padova, e\\
Istituto Nazionale di Fisica Nucleare, sezione di Padova, \\
via Marzolo 8, Padova I-35131, Italia,}
\affiliation{
$^{(2)}$Department of Physics and Astronomy,
University of Manitoba,
and Winnipeg Institute for Theoretical Physics,
Winnipeg, Manitoba, Canada R3T 2N2,}
\affiliation{
$^{(3)}$School of Physics, University of Melbourne,
Victoria 3010, Australia,}
\affiliation{
$^{(4)}$Advanced Research Computing, Information Division,
University of Melbourne, Victoria 3010, Australia.}

\begin{abstract}
A multi-channel algebraic scattering theory has been used to study the
properties of nucleon scattering from ${}^{12}$C and of the sub-threshold
compound nuclear states, accounting for properties in the compound nuclei to 
$\sim$ 10 MeV.  All compound and quasi-compound resonances observed 
in total cross-section data are matched, and on seeking solutions of the method
at negative energies, all sub-threshold states in ${}^{13}$C and ${}^{13}$N
are predicted with the correct spin-parities and with reasonable values
for their energies. A collective-model prescription has been used to define
the initiating nucleon-${}^{12}$C interactions and via use of orthogonalizing
pseudo-potentials, account is made of the Pauli principle.
Information is extracted on the underlying structure of each state in 
the compound systems by investigating the zero-deformation limit of the 
results.
\end{abstract}
\date{\today}
\maketitle

\section{Introduction}
\label{Intro}

In a recent paper~\cite{Am03}, a multichannel algebraic scattering (MCAS)
theory for nucleons scattering from a nucleus was specified in detail. At low
energies, the 
approach is noteworthy because its formulation facilitates a systematic 
determination both of the sub-threshold  bound states and of the compound 
resonances.  The theory is built upon sturmian expansions of whatever one 
chooses to be an interaction matrix of potential functions~\cite{Ca88}.  Of 
course there is the usual limitation of dealing with a coupled channels 
problem namely that the dimension of the evaluation rapidly increases with 
the number of channels considered. However, the MCAS approach does treat all
selected channels equivalently, whether they be open or closed, so that 
solutions can be found at both positive (scattering) and negative 
(bound system) energies relative to the nucleon on nucleus threshold.

While the MCAS approach may be used for any target (and projectile)
system, our formulation to date has been for nucleon-nucleus interactions.
However, the most practical adaptations insofar as size of the problem
and, concomitantly, computation times, are with light mass targets.
They have well separated low excitation spectra and usually the total
cross-section data show distinct resonances upon a smooth background.
Thus the first MCAS study~\cite{Am03} was of low-energy neutron-${}^{12}$C
(n+$^{12}$C) scattering. For that only three states of ${}^{12}$C ostensibly 
were needed in the evaluations; namely the $0^+_1$ (ground), $2^+_1$ (4.4389 
MeV), and the $0^+_2$ (7.6542 MeV). Herein we consider that same system in more
detail and, as well, analyze proton-${}^{12}$C (p+$^{12}$C) data.

The MCAS method solves the coupled-channel Lippmann-Schwinger equations
for the nucleon-nucleus system considered.  The
starting matrix of potentials may be constructed from any nuclear model: shell
or collective, rotational or vibrational, model.
We have used a rotational collective-model representation for
the interaction potential matrix but therein take deformation to second order.
We have chosen Woods-Saxon functions and their various derivatives to
be the form factors for all components each with characteristic
operators of central,
spin-orbit ($\mathbf{l} \cdot \mathbf{s}$), orbit-orbit ($\mathbf{l} \cdot 
\mathbf{l}$),
and of spin-spin ($\mathbf{s} \cdot \mathbf{I}$) type. The interactions were
allowed to depend on parity as well.  With such a characterization, the model
potential is sufficiently flexible to describe all possible structures (at
both positive and negative energies) of the nucleon-${}^{12}$C system.
However any such collective model prescription violates the Pauli principle.
At these energies that violation is severe.
But it is possible~\cite{Am03,Ca04} in the
MCAS approach to account for Pauli blocking of occupied nucleon
states in the target, with this or any collective model specification of the
matrix of potentials.  That is achieved by introducing orthogonalizing
pseudo-potentials~\cite{Ku78,Sa69,Ca04} (OPP) into the scheme by which the
sturmians are specified.  In that way all sturmians in the (finite) set selected as the
basis of expansion of the matrix of potentials contain
few or no components equivalent to the external nucleon being
placed in an already densely occupied orbit. Treatment of the
Pauli principle has a significant effect on results~\cite{Am03,Ca04}.

The aim of this paper is to extend calculations from the $^{13}$C to the 
$^{13}$N system, and to carry out a comparative analysis of bound and resonance
spectra involved.
A second version of the program has been set up, able to deal on the same
footing with the scattering of neutrons and protons from the target nucleus.
The neutron results confirm those of Ref.~\cite{Am03}.

It will be shown that, in the n+$^{12}$C process, the spectrum of resonances
up to about 6 MeV (in the laboratory (lab) system) is almost completely 
described by a mechanism involving excitation of the first $2^+$ level of 
$^{12}$C with energy $\epsilon_2 = 4.4389$ MeV. The spectrum shows a sequence 
of compound 
resonances, generated by the $\frac{1}{2}^-$, $\frac{1}{2}^+$, and 
$\frac{5}{2}^+$ bound (single nucleon) states in $^{13}$C.
The situation is very similar in the p+$^{12}$C process, with one overall
energy shift due to the Coulomb interaction. Because of this shift in energy,
some of the compound resonances in the n+${}^{12}$C system become 
quasi-compound ones in the p+${}^{12}$C case.

We will show that when 
deformation (which for the nucleon-${}^{12}$C system we consider is 
of quadrupole type and so linked to a parameter $\beta_2$) tends to zero, the
compound and quasi-compound states tend to pure states. For compound
resonances, their widths tend to zero and their centroid energies tend to those
of single particle bound states plus the core excitations, $\epsilon_i$. 
For quasi-compound resonances, the widths tend to the natural widths of the 
single particle resonance upon which they are formed, while their centroid 
energies tend to those of the single particle resonance plus the core 
excitations $\epsilon_i$.
It is interesting then to analyze the behavior of the phenomenology
contained in the model as $\beta_2$ varies continuously from the deemed
physical value to zero. There is a double purpose to this, namely to check the 
rules outlined above in a significant physical case and to describe the 
spectroscopy of $^{13}$C and $^{13}$N in terms of the scheme specified.

In Sec.~\ref{Sec2}, we discuss the results on the n+$^{12}$C system,
some of which were reported previously~\cite{Am03}. But now we analyze the 
origin of each state by studying results in the zero deformation limit.
In Sec.~\ref{Sec3}, the same analysis is extended to a study of the p+$^{12}$C 
system, where the Coulomb shift transforms some of the compound resonances into 
quasi-compound ones.  All calculations of this system have been carried out 
assuming charge symmetry. The potentials that gave the results discussed in
Sec.~\ref{Sec2} (and previously~\cite{Am03}) have been used without change. We 
present conclusions we have drawn from these studies in Sec.~\ref{Sec4}.

\section{The $^{13}$C system}
\label{Sec2}

Calculations of the n+$^{12}$C system have been carried out with the 
parameters defining the initiating nucleon-nucleus interaction matrix of
potentials being those used previously~\cite{Am03}. They are presented again in 
Table~\ref{Tab1} for easy reference.
\begin{table}
\caption{\label{Tab1} n+$^{12}$C potential parameters
(strengths in MeV).}
\begin{ruledtabular}
\begin{tabular}{ccccc}
parity & central & orbit-orbit & spin-orbit & spin-spin \\ 
\hline
$-$ & $-$49.1437 & 4.5588 & 7.3836 & $-$4.7700 \\
+ & $-$47.5627 & 0.6098 & 9.1760 & $-$0.0520 \\
\hline
other parameters: & & & \\
& $r_0 = 1.35$ fm. & $a_0 = 0.65$ fm. & $\beta_2 = -0.52$
\end{tabular}
\end{ruledtabular}
\end{table}
Couplings of the input channel with the ground state ($\epsilon_1 = 0$), with 
the $2^+$ ($\epsilon_2 = 4.4389$ MeV) and with the $0^+_2$ ($\epsilon_3 =
7.6542$ MeV) excited states of the $^{12}$C target have been considered. 
Moreover, the Pauli principle has been taken into account throughout using the
OPP procedure~\cite{Am03,Ca04}. However, note that the OPP treatment~\cite{Am03} 
we use is an approximation. We have used a large but finite strength in place
of the infinite value that the OPP theory presumes.  One must thus expect some
small spuriosity, which may affect our calculated energies but by 
about a tenth of an MeV.

\begin{figure}
\scalebox{0.75}{\includegraphics*{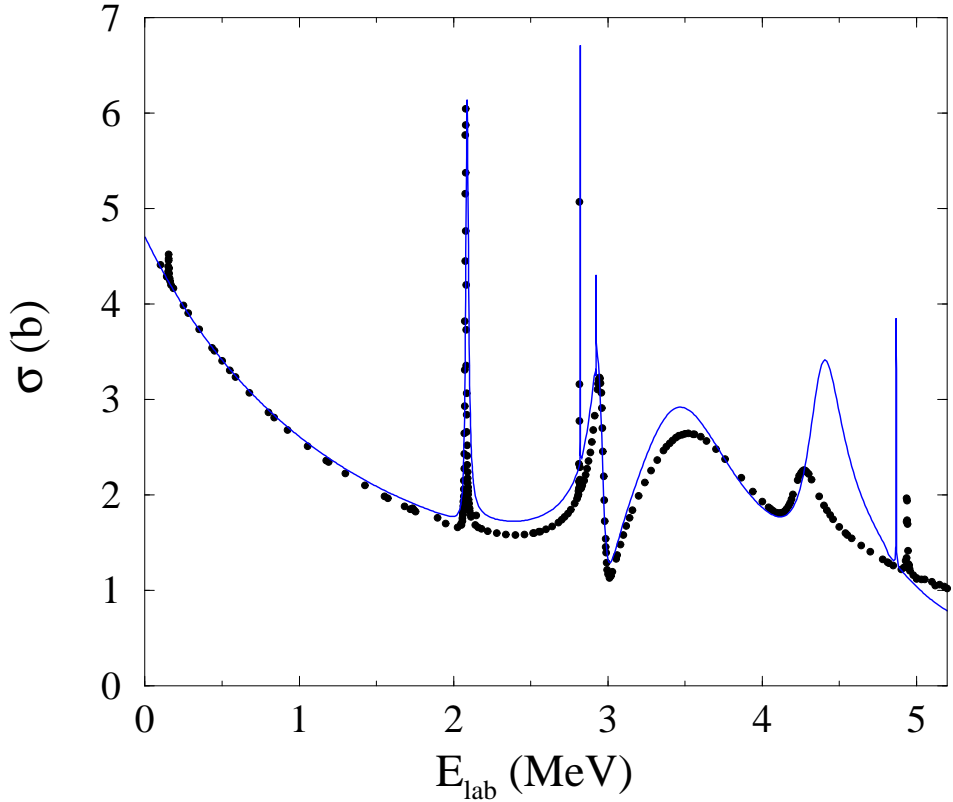}}
\caption{\label{Fig1}(Color online)
Comparison between data (filled circles) and theoretical (solid curve)
calculated elastic scattering cross sections for n+${}^{12}$C scattering
to $\sim 5$ MeV.}
\end{figure}

In Fig.~\ref{Fig1}, our calculated n+$^{12}$C elastic cross section is compared
with the evaluated nuclear data file (ENDF) formed by Pearlman~\cite{Pe93}.
Source data and references were obtained using the computer index of neutron data 
(CINDA)~\cite{IA05}.
The energies are in the laboratory frame.
The sequence of resonances evident therein as energy increases have spin-parity
assignments $\frac{5}{2}^+$, $\frac{7}{2}^+$, $\frac{1}{2}^-$, $\frac{3}{2}^+$,
$\frac{3}{2}^+$, $\frac{5}{2}^+$, $\frac{9}{2}^+$. Details of those resonances
and of bound states are compared with experimental values~\cite{Aj91} in
Table~\ref{Tab2}. We have found the  parameters of the resonances by studying 
the trajectories of the sturmian eigenvalues using the algorithm defined in 
Ref.~\cite{Am03}. This we defined as the resonance identification (RI) process.
As already noticed in Ref.~\cite{Am03}, this choice is unambiguous in the case 
of narrow resonances, while for wider resonances it represents just one of the
possible methods for resonances identification. Note that in this table, the 
energies of the states (in MeV) are given in the center of mass (cm) system. 
The widths, on the other hand, are expressed in keV.
\begin{table}
\caption{\label{Tab2} Sub-threshold bound states and low 
energy resonances of the n+$^{12}$C system.}
\begin{ruledtabular}
\begin{tabular}{cccccc}
entry (n) & $J^ \pi$ & $E_{\text{exp}}$ & $\frac{1}{2}\Gamma_{\text{exp}}$(keV) 
& $E_{\text{th}}$ & $\frac{1}{2}\Gamma_{\text{th}}$(keV) \\
\hline
1 & $\frac{1}{2}^-$ & $-$4.9463 & --- & $-$4.8881 & --- \\
2 & $\frac{1}{2}^-$ & --- & --- & 2.6829 & 0.332 \\
3 & $\frac{1}{2}^+$ & $-$1.8569 & --- & $-$2.0718 & --- \\
4 & $\frac{1}{2}^+$ & --- & --- & 4.6629 & 555 \\
5 & $\frac{3}{2}^-$ & $-$1.2618 & --- & $-$1.4783 & --- \\
6 & $\frac{3}{2}^+$ & 2.7397 & 35 & 2.7309 & 40.8 \\
7 & $\frac{3}{2}^+$ & 3.2537 & 500 & 3.2447 & 447 \\
8 & $\frac{5}{2}^-$ & 0.1 & --- & $-$0.0338 & --- \\
9 & $\frac{5}{2}^+$ & $-$1.0925 & --- & $-$1.8619 & --- \\
10 & $\frac{5}{2}^+$ & 1.9177 & 3 & 1.9348 & 9.65 \\
11 & $\frac{5}{2}^+$ & 3.9314 & 55 & 4.0579 & 126 \\
12 & $\frac{7}{2}^+$ & 2.547 & $\leq$2.5 & 2.6220 & 8.74 $\times 10^{-4}$ \\
13 & $\frac{9}{2}^+$ & 4.534 & 2.5 & 4.5091 & 0.745
\end{tabular}
\end{ruledtabular}
\end{table}
In the first column of Table~\ref{Tab2} we give an identifying number (n) 
to each state which we use throughout the following discussion.  

In the spectrum, the 
$\frac{1}{2}^-$ (n~=~2) resonance is not seen experimentally. However, from our 
model calculation it is practically coincident with the first and strong
$\frac{3}{2}^+$ resonance (n~=~6), and is shown in the figure like a narrow spike
over the broader $\frac{3}{2}^+$ peak. Possibly that is why it has not seen in
experiments to date.  But an isobaric analogue to this has been observed
in $^{13}$N, and its nature is well explained within the general 
logic of the spectroscopic structure we believe to be sensible.
The very narrow $\frac{5}{2}^-$ resonance (n~=~8) which lies just above threshold 
has a partner in our calculations, but as a bound state just below threshold.

\begin{figure}
\scalebox{0.75}{\includegraphics*{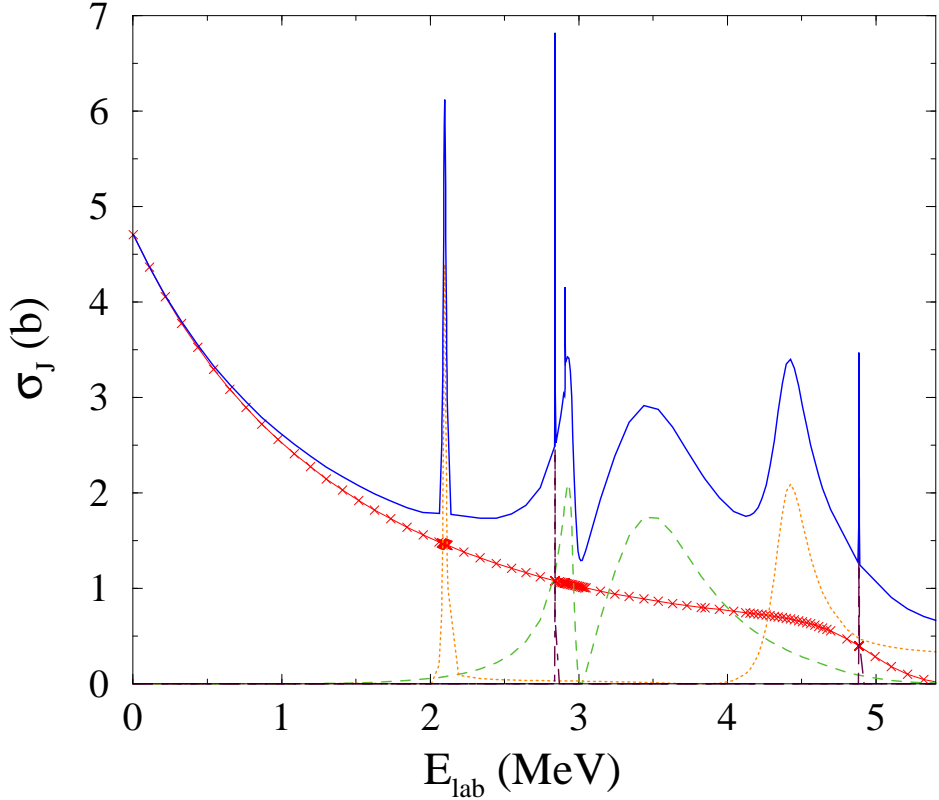}}
\caption{\label{Fig2}(Color online)
Components in the n+$^{12}$C elastic cross section. The theoretical curve 
(solid curve) is decomposed in its even-parity components. The $\frac{1}{2}^+$ 
component is shown by the generally smooth curve connecting `$\times$' marks.
It constitutes virtually the entire cross section at energies near threshold.  
The $\frac{3}{2}^+$ and  $\frac{5}{2}^+$ components are portrayed by the dashed and
dotted curves respectively. The narrow $\frac{7}{2}^+$ and $\frac{9}{2}^+$ resonances
at (lab) energies of 2.84  and 4.88 MeV respectively are portrayed by long dashed
curves.}
\end{figure}

The peak shown in Fig.~\ref{Fig1} at the energy of about 4.40 MeV in the laboratory 
frame (4.06 MeV in the cm system) corresponds to our calculated $\frac{5}{2}^+$ 
(n~=~11) resonance. In the neighborhood the model also predicts (by using the  
RI procedure) a $\frac{1}{2}^+$ resonance (n~=~4) whose centroid
energy is $E_{\text{lab}} = 5.05$ MeV ($E_{\text{cm}} = 4.66$ MeV).  No 
corresponding peak
in the elastic cross section for this is seen. In fact, that resonance is 
missing entirely in Fig.~\ref{Fig1}.  To solve this puzzle we calculated the 
even-parity components of the cross section separately; i.e. the cross section 
was calculated for the unique $J^\pi$ value of $\frac{1}{2}^+$, then with 
$\frac{3}{2}^+$, etc.  The results shown in Fig.~\ref{Fig2} clearly identify
the component responsible for each even-parity resonance. The only peak not 
accounted for in the analysis is that having a spin-parity $\frac{1}{2}^-$.
In particular under
the maximum at about 4.4 MeV, there are two overlapping resonances. 
They are the $\frac{5}{2}^+$ component which has a sharp maximum in the energy 
region under consideration and the $\frac{1}{2}^+$ component revealed as a very 
small bump at about the same energy.  It is a very broad resonance. The 
$\frac{1}{2}^+$ component is more easily recognized in the calculated
cross section when, as we discuss later, the deformation is decreased.  
Thus the cross section is, as could be expected, a dominantly $s$-wave 
background with strong narrow and broad resonances superimposed. The individual
values of energy at which the $s$-wave cross section initially was calculated 
are represented by the $\times$ in the plot for $J^\pi = \frac{1}{2}^+$.
The clustering of $\times$'s (around regions in which sharp resonances in the total
cross section occur) reflects the density of mesh points used 
to find the variation through those regions as precisely as possible. 
Such requirement is further evidence of the need in any such study to ensure 
that all resonance centroids and half widths are defined in the process, 
and energy steps in that region are selected appropriately, otherwise with too 
large an energy step they will be missed.  The RI procedure~\cite{Am03} 
ensures all resonances will be found in the overall energy regime to be studied. 
Also built into the code 
implementing this procedure is an automatic increase in numbers of energy 
points. The span of those points is dependent upon the half width 
of the resonance, as defined by the RI procedure.

\begin{figure}
\scalebox{0.75}{\includegraphics*{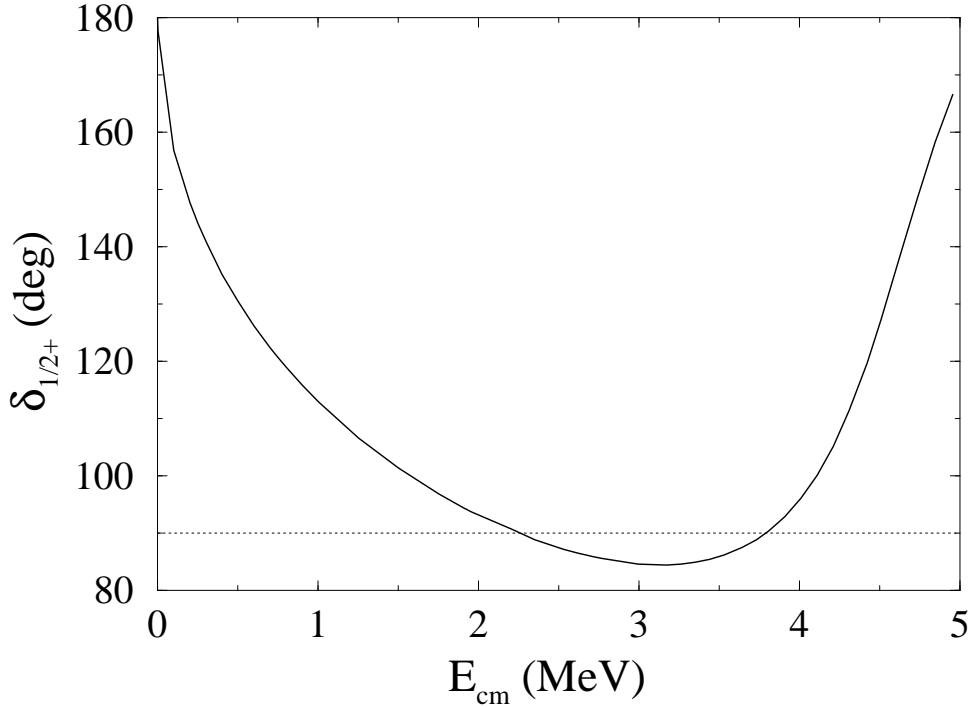}}
\caption{\label{Fig3}
Energy variation of the calculated $\frac{1}{2}^+$ scattering phase shift.}
\end{figure}

We conclude therefore that the 4.4 MeV peak in the measured cross section is 
mainly a $\frac{5}{2}^+$ resonance. The $\frac{1}{2}^+$ resonance foreseen by 
the model calculations is weak and completely masked by it.  The 
effect may be understood better by studying the (cm) energy variation of the 
$\frac{1}{2}^+$ (total) scattering phase shift. That variation is displayed in 
Fig.~\ref{Fig3}. 
This phase shift does not exhibit a regular increment between $0^\circ$ and $180^\circ$
(as do the other resonant phase shifts), but start from $180^\circ$ (because of 
Levinson's theorem) at zero energy, reaching a minimum of $84.5^\circ$ at 3.17~MeV 
and then increases passing through  $173^\circ$ at 5.063MeV. The
result is a broad and not very pronounced peak in the partial cross section, as
shown in Fig.~\ref{Fig2}. Hence we assign our $\frac{5}{2}^+$ (n~=~11) level 
with the well defined experimental resonance at $E_{\text{cm}} = 3.9314$ MeV 
rather than as the $\frac{1}{2}^+$ state heretofore assumed in the literature~\cite{Aj91}.

The agreement between theory and experiment shown in Fig.~\ref{Fig1} and in
Table~\ref{Tab2} is very good. Energies of both bound and resonance states, and
the widths of those resonances, are quite well matched. Particularly so when 
one
recalls that the target structure has been taken within a simple rotational
model scheme. A measure of the goodness of fit is the mean square error
in the table entries.  That measure formed using 11 of the states is
\begin{equation}
\label{msqE}
\mu= \frac{1}{N} \sqrt{\sum (E_{\text{th}} - E_{\text{exp}})^2 }=0.0776 \; 
{\rm MeV}.
\end{equation}
Entries n~=~2 and n~=~4 have been omitted because the first has been not detected
experimentally while for the second, we propose a different assignment of 
quantum numbers.

Our procedure enables us to interpret the structure of the spectra by 
following the general discussion that has been given~\cite{Pi95} about the 
properties of compound resonances.
Consider first the even-parity states. These can be specified in the chosen 
representation in terms of pure states that are identified by the set of 
quantum numbers, \{$J^\pi, I, j, \ell$\}.  While the relevant details 
concerning these sets of quantum numbers have been extensively discussed
before~\cite{Am03}, it is useful to note here that
the single particle quantum numbers ($l$) and ($j$) can be
related to a state of the single nucleon in the mass-13 systems. Then, in the 
energy range we study, and considering only the effect of the most important 
couplings, i.e. with the ground and $2^+$ state of $^{12}$C, all possible 
configurations are as listed in Table~\ref{Tab3}. The second excited $0^+_2$
state will enter discussion later in relation to a particular ($\frac{1}{2}^-$)
excited state in the mass-13 spectra. 
\begin{table}
\caption{\label{Tab3} Quantum numbers of allowed even-parity basis states
($J^{\pi} \leq \frac{9}{2}^+$).}
\begin{ruledtabular}
\begin{tabular}{ccccccccccc}
N & $J^\pi$ &  I & j & l & \vline & N & $J^\pi$ & I & j & l \\ 
\hline
1 & $\frac{1}{2}^+$ & 0 & $\frac{1}{2}$ & 0 & \vline & 14 & $\frac{5}{2}^+$ & 2 &
$\frac{9}{2}$ & 4 \\
2 & $\frac{1}{2}^+$ & 2 & $\frac{3}{2}$ & 2 & \vline & 15 & $\frac{7}{2}^+$ & 0 &
$\frac{7}{2}$ & 4 \\
3 & $\frac{1}{2}^+$ & 2 & $\frac{5}{2}$ & 2 & \vline & 16 & $\frac{7}{2}^+$ & 2 &
$\frac{3}{2}$ & 2 \\
4 & $\frac{3}{2}^+$ & 0 & $\frac{3}{2}$ & 2 & \vline & 17 & $\frac{7}{2}^+$ & 2 &
$\frac{5}{2}$ & 2 \\
5 & $\frac{3}{2}^+$ & 2 & $\frac{1}{2}$ & 0 & \vline & 18 & $\frac{7}{2}^+$ & 2 &
$\frac{7}{2}$ & 4 \\
6 & $\frac{3}{2}^+$ & 2 & $\frac{3}{2}$ & 2 & \vline & 19 & $\frac{7}{2}^+$ & 2 &
$\frac{9}{2}$ & 4 \\
7 & $\frac{3}{2}^+$ & 2 & $\frac{5}{2}$ & 2 & \vline & 20 & $\frac{7}{2}^+$ & 2 &
$\frac{11}{2}$ & 6 \\
8 & $\frac{3}{2}^+$ & 2 & $\frac{7}{2}$ & 4 & \vline & 21 & $\frac{9}{2}^+$ & 0 &
$\frac{9}{2}$ & 4 \\
9 & $\frac{5}{2}^+$ & 0 & $\frac{5}{2}$ & 2 & \vline & 22 & $\frac{9}{2}^+$ & 2 &
$\frac{5}{2}$ & 2 \\
10 & $\frac{5}{2}^+$ & 2 & $\frac{1}{2}$ & 0 & \vline & 23 & $\frac{9}{2}^+$ & 2 &
$\frac{7}{2}$ & 4 \\
11 & $\frac{5}{2}^+$ & 2 & $\frac{3}{2}$ & 2 & \vline & 24 & $\frac{9}{2}^+$ & 2 &
$\frac{9}{2}$ & 4 \\
12 & $\frac{5}{2}^+$ & 2 & $\frac{5}{2}$ & 2 & \vline & 25 & $\frac{9}{2}^+$ & 2 &
$\frac{11}{2}$ & 6 \\
13 & $\frac{5}{2}^+$ & 2 & $\frac{7}{2}$ & 4 & \vline & 26 & $\frac{9}{2}^+$ & 2 &
$\frac{13}{2}$ & 6
\end{tabular}
\end{ruledtabular}
\end{table}
Assuming a shell model single nucleon spectrum for an additional nucleon
coupling to states in $^{12}$C we expect at most three even-parity bound states 
to be 
important, namely relating to lodgment of that nucleon within the $1s$-$0d$ 
shell.  However, as reported~\cite{Am03}, shell model 
studies suggest that the low excitation mass-13 states dominantly 
are identifiable with either
$0d_\frac{5}{2}$ or $1s_\frac{1}{2}$ orbit couplings to the ground and 
$2^+$ states in ${}^{12}$C.  As such, the coupling to the ground state 
(in ${}^{12}$C) will provide two states in the compound nucleus with 
spin-parity values of  $\frac{1}{2}^+$ and $\frac{5}{2}^+$. Therefore we expect
one doublet and one quintuplet of even-parity states when the coupling is made 
with the $2^+$ state. 

Using this scheme, the bound state $\frac{1}{2}^+$ (n~=~3 of Table~\ref{Tab2}) 
in ${}^{13}$C should be dominantly described as a $1s_{\frac{1}{2}}$ 
neutron bound to the $^{12}$C ($0^+$) ground state core and identified with 
the N = 1 entry in the table of pure states: this infers a bound state of 
energy 
$E \simeq -2$~MeV. Likewise there should be a $\frac{5}{2}^+$ state (n=9
of Table~\ref{Tab2}), which can be
identified with the N = 9 entry in Table~\ref{Tab3}.  Then, when a neutron 
impinging with energy $\epsilon_2 + E$ loses $\epsilon_2$ to
excitation and is bound to the $^{12}$C$^*~(2^+)$ core, there should arise a 
very narrow resonance; a bound state in the continuum for zero coupling. As the 
coupling increases, this resonance splits forming a doublet for which the  
quantum numbers will be  $J^{\pi} = \frac{3}{2}^+$ and $\frac{5}{2}^+$.  The 
first of these we identify with the state N = 5 (responsible for the resonant 
behavior), coupled with an elastic background arising from the effect of the 
N = 4 entry.  The second of the doublet we link to the state N~=~10 and expect 
that it is attached to a background state for which N = 9. As the deformation 
($\left| \beta_2 \right|$) increases, the splitting of the doublet, and the resonance 
widths, 
increase. Also all of the $J$-components contribute since entries  N=4, 5, 6, 7,
and 8 are involved for $J = \frac{3}{2}$ and those of  N=9, 10, 11, 12, 13, 
and 14 are so for $J = \frac{5}{2}$.  Hence the bound state 
$\frac{1}{2}^+$ is expected to generate a doublet of compound resonances, and 
by the same mechanism, the bound state $\frac{5}{2}^+$ when coupled to the 
$2^+$ state should generate a quintuplet with spin-parities $J= \frac{1}{2}^+, 
\frac{3}{2}^+ , \frac{5}{2}^+, \frac{7}{2}^+,\ {\rm and}\ \frac{9}{2}^+$.

\begin{figure}
\scalebox{0.7}{\includegraphics*[clip]{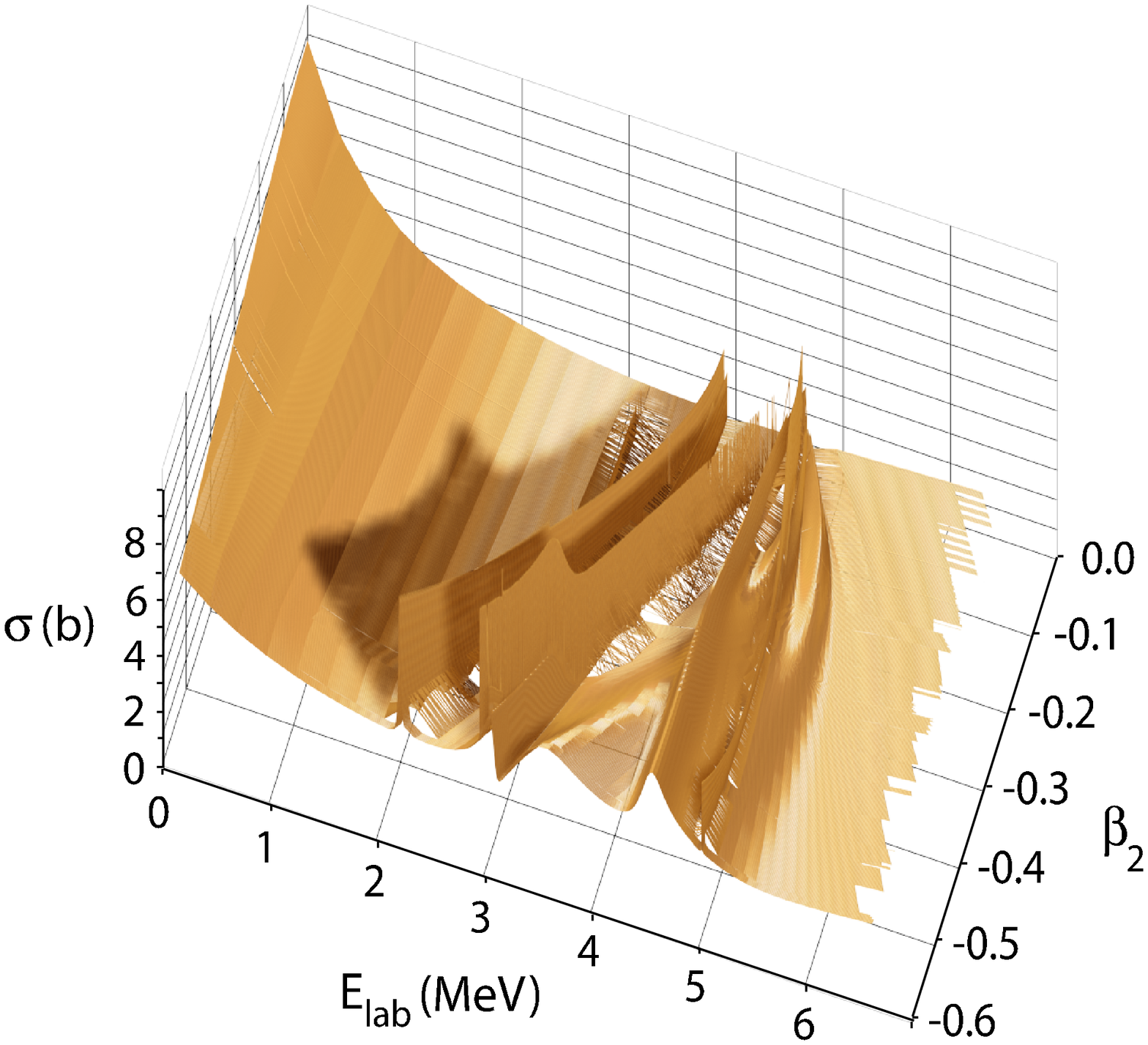}}
\caption{\label{Fig4}(Color online)
Total elastic cross sections for n+$^{12}$C scattering as functions
of neutron energy showing the effects of reducing the value of
$\beta_2$. Details are as described in the text.}
\end{figure}

To verify this scheme, we have made calculations starting with the physical 
situation  of having $\beta_2 = -0.52$, and then form cross sections from
calculations in which the coupling is gradually reduced. By that means
it is possible to track each state/resonance continuously and so
identify the underlying base  nature of each state. It is assumed that each
resonance conserves its
identity in the adiabatic transition from the physical to the unperturbed 
limit.  We portray the results in Fig~\ref{Fig4},
which shows the general trend that the background cross-section values near threshold 
increase in size as $\left| \beta_2 \right|$ decreases. That is consistent with the 
strong sub-threshold $s$-wave ${\frac{1}{2}}^+$ state moving closer to threshold as 
$\left| \beta_2 \right|$ decreases. Then what was just a sub-threshold state, the 
${\frac{5}{2}}^-$ in the physical case calculation, moves into the positive 
energy regime with decrease in $\left| \beta_2 \right|$. That state has an extremely small 
width ($\le 10^{-11}$ MeV) so that while its existence is known, its strength 
has not been ascertained.  The actual resonances also have unique trends with 
decreasing $\left| \beta_2 \right|$. With their identification (the entry `n' specified 
in Table~\ref{Tab2}), the narrow $\frac{5}{2}^+$ (n~=~10) and the first 
$\frac{3}{2}^+$ (n~=~6) track smoothly with decreasing width till they
have essentially the same centroid energy and vanishing width; features that
we discuss subsequently.  The very narrow $\frac{7}{2}^+$ (n~=~12) resonance
initially sits upon the $\frac{3}{2}^+$ resonance shape and its centroid 
increases
in value with decreasing $\left| \beta_2 \right|$. It remains a feature upon 
the 
$\frac{3}{2}^+$ shape until $\left| \beta_2 \right| \preceq 0.35$, after which 
that 
$\frac{7}{2}^+$ centroid moves 
up in energy to be one of the quintuplet of sharp resonances at $\sim 4.3$ MeV.
The 
other four members of the quintuplet are the very broad second $\frac{3}{2}^+$ 
(n~=~7), the reasonably broad strong $\frac{5}{2}^+$ (n~=~11), the sharp 
$\frac{9}{2}$ 
(n~=~13), and the reasonably broad but weak $\frac{1}{2}^+$ ($\sim$4.6 MeV) 
resonances.
The last is not readily seen until, with decreasing $\left| \beta_2 \right|$, 
the width of 
the nearby $\frac{5}{2}^+$ resonance decreases sufficiently. In the figure, 
this 
$\frac{1}{2}^+$ resonance is seen as the highest energy track that curves back 
in energy to meet
the rest of the quintuplet when $\beta_2 = 0$. The $\frac{9}{2}^+$ resonance 
sits 
initially on the tail of the $\frac{5}{2}^+$ resonance appearing to gradually 
decrease 
in size ( because the $\frac{5}{2}^+$ resonance is contracting as $\left| 
\beta_2 \right|$ 
decreases) before some enhancement at small values of $\left| \beta_2 \right|$ 
due, then, to the proximity of all members of the quintuplet. The centroids of the two broad 
resonances (the $\frac{3}{2}^+$ and the $\frac{5}{2}^+$) of this quintuplet cross as
$\left| \beta_2 \right|$ decreases and while each width gradually decreases, they combine 
to form quite distinctively changing cross sections until, as $\beta_2 \to 0$, they 
become very narrow and tend to the same centroid energy.  
Of course when $\beta_2$ is exactly zero, since the widths are vanishingly small, a
calculated cross section loses almost all trace of individual compound 
resonances. Nevertheless by means of the RI procedure, the 
resonance centroid energies may still be (and were) found. 

\begin{figure}
\scalebox{0.75}{\includegraphics*{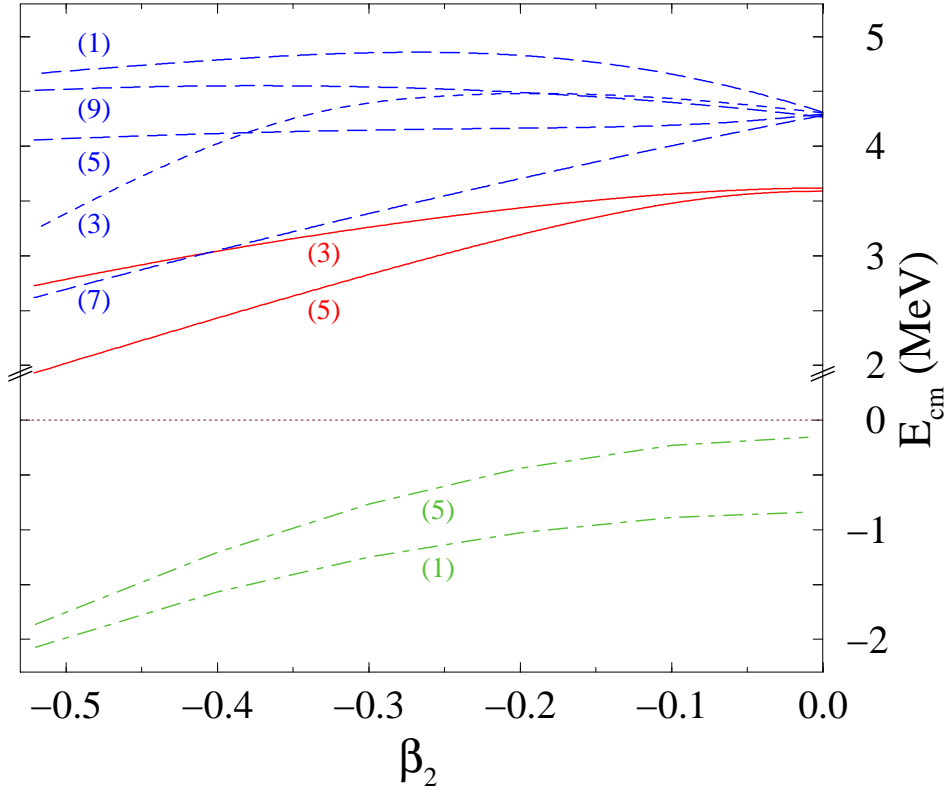}}
\caption{\label{Fig5}(Color online)
Variation of energies of the $\frac{1}{2}^+$ bound and doublet, and of the 
$\frac{5}{2}^+$ 
bound and quintuplet of states with deformation. The numbers in brackets 
attached to each curve are the specific values of $2J$.} 
\end{figure}

In Fig.~\ref{Fig5} we show our results of the calculations of the centroids of
the positive parity resonances and sub-threshold bound states.
Consider first the states built upon a $\frac{5}{2}^+$ single particle state.
They are the quintuplet of resonances having spin-parities
$J^\pi = \frac{1}{2}^+$, $\frac{3}{2}^+$, $\frac{5}{2}^+$, $\frac{7}{2}^+$,
\ {\rm and}\ $\frac{9}{2}^+$ plus the sub-threshold (bound) state in ${}^{13}$C.  
Their energy values when $\beta_2 = 0$ (i.e. $\sim -0.15$ and 4.28) equate to the 
target excitation $4.28 - (-0.15) = 4.43 \simeq \epsilon_2$, and so we conclude that 
the quintuplet indeed is a set of compound resonances due to coupling of a 
$\frac{5}{2}^+$ 
neutron with the $2^+$ excited state of $^{12}$C and that the $0d_{\frac{5}{2}}$
nucleon binding to the ground state of ${}^{12}$C is $\sim -0.15$ MeV. In a similar
way, the $\frac{1}{2}^+$ bound state (n~=~3) has a binding to the
ground state of ${}^{12}$C of $\sim -0.83$ MeV and the doublet of resonances 
(n~=~6 and 10) relate in energy to identify
their origins as the $\frac{3}{2}^+$, $\frac{5}{2}^+$ doublet of states
generated by that $\frac{1}{2}^+$ neutron being coupled to the $2^+$ excited 
state in ${}^{12}$C. In the zero deformation limit, the energies of the bound 
and collapsed doublet
states are $\sim -0.83$ and 3.60 MeV thereby differing by the 4.43 
$\simeq \epsilon_2$ value.

This leads to a complete understanding of the spectroscopy of the even-parity
states in ${}^{13}$C as they are revealed by low energy n+${}^{12}$C scattering.
Note that the two  $\frac{3}{2}^+$ states (n~=~6 and 7) that come from analysis 
solely of the scattering data, basically are indistinguishable because only 
$J^\pi$ is conserved.  Nevertheless we expect that n~=~6 is part of the doublet 
and n~=~7 is part of the quintuplet because of the continuity in the limit 
$\beta_2 \rightarrow 0$.
Of course, similar information should be obtained by large-space structure 
model analyses of the wave functions in the two cases. Work is in progress to 
do just that.

We note that the $\beta_2 = 0$ limit values of energies of the even-parity 
states discussed tend only approximately to the same values. There are
small but significant shifts.  That effect is due to the residual operator
character of the interaction potentials.  While the coupling between the
channels due to deformation was removed, there remains some linkages
arising from the diagonal potentials being
\begin{equation}
V_c(r) \equiv V_{cc}(r)=\left[V_0 + \ell(\ell+1)V_{\ell\ell}\right]f_0(r)
- \frac{1}{ar} W_{ls} A(r) \left[{\bf l} \cdot {\bf s}\right]_{cc} 
+ V_{ss} f_0(r)  \left[{\bf s} \cdot {\bf I}\right]_{cc} .
\label{Vbeta-0}
\end{equation}
This is the reduced form of Eq.~(B.5) of Ref~\cite{Am03}, in the limit
$\beta_2\rightarrow 0$, where it is meant that the channel index 
$c$ denotes the set of quantum numbers $\{J^\pi, I, j, \ell \}$.
Clearly the above discussion fails to hold exactly in presence of a 
spin-spin interaction. How much so is illustrated by the entries in Table~\ref{Tab5}.
\begin{table}
\caption{\label{Tab5} Properties of the even-parity
quintuplet of resonances in the limit $\beta_2 = 0$.}
\begin{ruledtabular}
\begin{tabular}{ccccccc}
n & $J^ \pi$ &  $E_{\text{th}}$ & $K(\ell, j, j, J$) & $\Delta E_{\text{th}}$ 
& $\Delta K$ & $\left| \frac{\Delta K}{\Delta E_{\text{th}}} \right|$ \\ \hline
4 & $\frac{1}{2}^+$ & 4.3094 & $-$14 & --- & --- & --- \\
7 & $\frac{3}{2}^+$ & 4.3039 & $-$11 & $-$0.0055 & 3 & 545 \\
11 & $\frac{5}{2}^+$ & 4.2948 & $-$6 & $-$0.0091 & 5 & 549 \\
12 & $\frac{7}{2}^+$ & 4.2820 & 1 & $-$0.0128 & 7 & 546 \\
13 & $\frac{9}{2}^+$ & 4.2656 & 10 & $-$0.0164 & 9 & 548
\end{tabular}
\end{ruledtabular}
\end{table}
In this table we have selected the states of the even-parity quintuplet, and
show in column 3, the calculated values of the unperturbed energies. They
reveal the small defects of convergence caused by the spin-spin (target) specific 
potential term. It is easy to recognize that
the members of the even-parity quintuplet (in the unperturbed conditions) are 
pure states of the type N = 3, 7, 12, 17, and 22, as given in Table~\ref{Tab3}
with quantum numbers $c \equiv \{\frac{1}{2}^+, 2, \frac{5}{2}, 2 \}$,
$\{\frac{3}{2}^+, 2, \frac{5}{2}, 2 \}$,
$\{\frac{5}{2}^+, 2, \frac{5}{2}, 2 \}$,
$\{\frac{7}{2}^+, 2, \frac{5}{2}, 2 \}$,\ {\rm and}\ 
$\{\frac{9}{2}^+, 2, \frac{5}{2}, 2 \}$ respectively. These states differ 
(only) in the total angular momentum $J$ and therefore are discriminated 
(only) by the spin-spin part of the potential.

In particular, from Eqs.~(B8), (B9) and (B10) of Ref.~\cite{Am03}, it may be
seen that the spin-spin potential eigenvalues depend on the quantum number $J$
through functions $K(\ell,j,j',J)$. Those numbers are listed in column 4 of 
Table~\ref{Tab5}.  To verify the proportionality of the values of $E_{\text{th}}$ 
and $K$, we give the differences $\Delta E_{\text{th}}$ between adjacent states in 
the list (i.e.  $E_{\text{th}}$(n~=~7) $-$ $E_{\text{th}}$(n~=~4), etc.) in column 5.  
Then, in column 6, the associated 
differences  $\Delta K$ (namely $K$(n~=~7) $-$ $K$(n~=~4), etc.) are given while in 
column 7 we present the ratios $\frac{\Delta K}{\Delta E_{\text{th}}}$.  The value 
reported in the 
last column is constant within 0.4 percent, most readily confirming the 
assumption that the degeneracy of the multiplet in the unperturbed limit is 
broken only by the spin-spin interaction. Therefore we have repeated calculations 
with $\beta_2 = 0$ but additionally with the spin-spin potential set to zero. 
All other parameters remain with those values given in Table~\ref{Tab1}.
The energies of the states then found are presented in Table~\ref{Tab6}. 
Therein also, in the rightmost column, we give the gap energies $\Delta E_n$
which are the 
differences between the resonance energies of the listed value and of the bound
state energy of the particle coupled to the $^{12}$C ground state.  Therein the 
states for n~=~1~,~3~,
\begin{table}
\caption{\label{Tab6} Sub-threshold bound states and resonances in 
n+$^{12}$C when $\beta_2 = 0$ and $V_{ss} = 0$.}
\begin{ruledtabular}
\begin{tabular}{cccc}
n & $J^ \pi$  & $E_{\text{th}}$ & $\Delta E_n$ \\ 
\hline
1 & $\frac{1}{2}^-$  & $-$4.7017 & 0 \\
2 & $\frac{1}{2}^-$  & 2.9525 & 7.6542 \\
3 & $\frac{1}{2}^+$  & $-$0.8376 & 0 \\
4 & $\frac{1}{2}^+$  & 4.2839 & 4.4389 \\
5 & $\frac{3}{2}^-$  & $-$0.2627 & 4.4390 \\
6 & $\frac{3}{2}^+$  & 3.6013 & 4.4389 \\
7 & $\frac{3}{2}^+$ & 4.2839  & 4.4389 \\
8 & $\frac{5}{2}^-$ & $-$0.2627 & 4.4390 \\
9 & $\frac{5}{2}^+$ & $-$0.1550 &  0 \\
10 & $\frac{5}{2}^+$  & 3.6013 & 4.4389 \\
11 & $\frac{5}{2}^+$  & 4.2839 & 4.4389 \\
12 & $\frac{7}{2}^+$ & 4.2839 & 4.4389 \\
13 & $\frac{9}{2}^+$ & 4.2839 & 4.4389
\end{tabular}
\end{ruledtabular}
\end{table}
and 9 are taken as coupling a $0p_\frac{1}{2}$, $1s_\frac{1}{2}$, and a
$0d_\frac{5}{2}$ neutron to the ground state of ${}^{12}$C. Then all of the 
remaining differences shown in the last column (except for the n~=~2 case)
are exactly the assumed excitation energy of the $2^+$ state in ${}^{12}$C
($\epsilon_2$).  As already observed, our assumption is that the even-parity 
quintuplet, n~=~4, 7, 11, 12, and 13, derives from the bound state giving
n~=~9 (and therefore the definitions of the gap energies are
$\Delta E_4=E_4-E_9$, $\Delta E_7=E_7-E_9$ etc), while
the even-parity doublet, n~=~6 and 10, derives from the bound state 
that lead to n~=~3 (and therefore $\Delta E_6 = E_6 - E_3$, $\Delta 
E_{10} = E_{10} - E_3$).  The n~=~2 entry results from the coupling of the 
$0p_\frac{1}{2}$ 
(that gave n~=~1) to the third state we chose to include in the target spectrum used 
in MCAS.  This is the only effect of the excitation of the $0^+_2$ excited
state in ${}^{12}$C at least for the energy range we have considered. 
To explore more effects of the $0^+_2$ state, higher energy results
need to be analyzed. But then one would also need consider effects of
other states such as the strong collective $3^-$ (9.63 MeV).

As far as the other  odd-parity states are concerned, the values listed in 
Table~\ref{Tab6} strongly suggest that the states n~=~5 and n~=~8 (which is 
reproduced as a near-threshold bound state in our calculations) are the result
of coupling a $0p_\frac{1}{2}$ neutron (as identified in the n~=~1 level)
with the $2^+$ excited state in ${}^{12}$C.  All these conclusions about the 
odd states may be drawn only after dropping the spin-spin term, because the 
spin-spin interaction is stronger for the odd states (see Table~\ref{Tab1}) and 
masks completely the effect of the convergence in the unperturbed limit.

\begin{figure}
\scalebox{0.75}{\includegraphics*{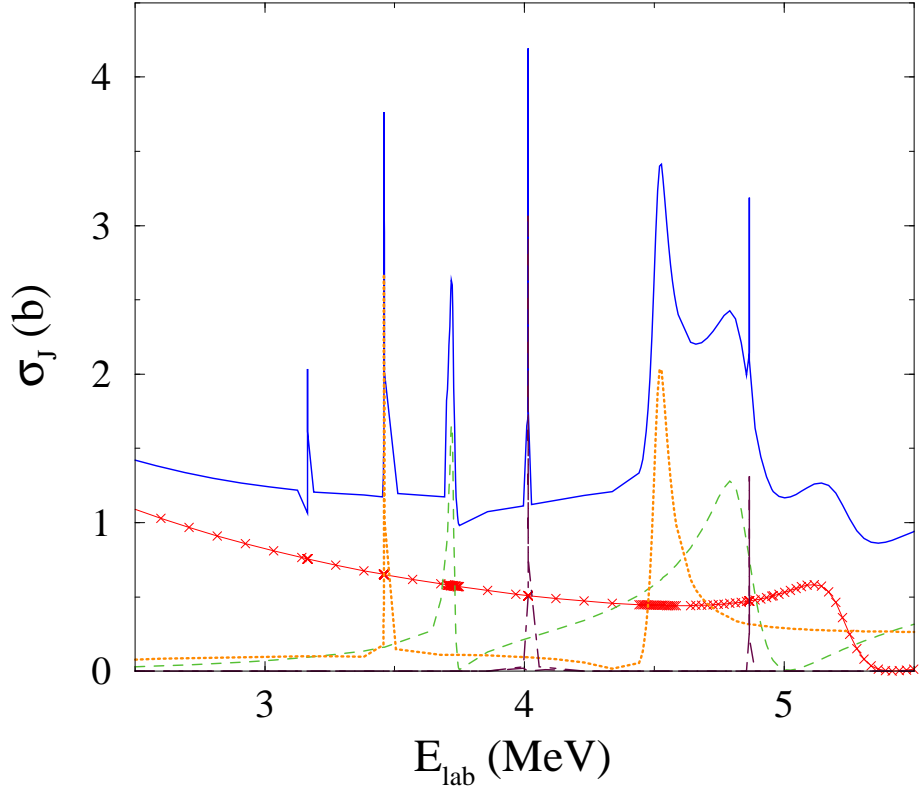}}
\caption{\label{Fig6}(Color online)
Even parity component contributions to the elastic n+$^{12}$C cross sections
in the case of small deformation, $\beta_2 = -0.2$. In this case the narrow
$\frac{7}{2}^+$ and $\frac{9}{2}^+$ resonances lie at (lab) energies of 4.0 and 
4.8 MeV.}
\end{figure}

\begin{figure}
\scalebox{0.75}{\includegraphics*{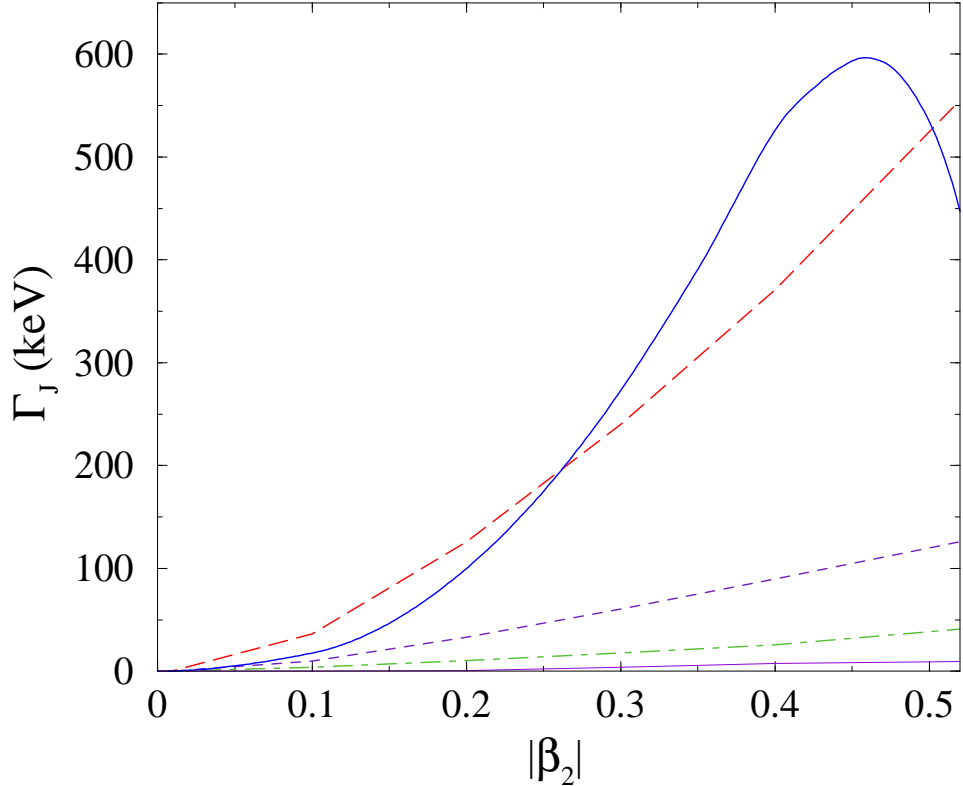}}
\caption{ \label{Fig7}(Color online)
Deformation dependence of widths, $\Gamma_{\text{th}} (J^\pi)$ (in keV), for some of 
the (broader) compound resonances. Those from cases with n~=~4, 6, 7, 10, and 11 are
portrayed by the long dashed, dot-dashed, solid, thin solid, and dashed
curves respectively.}
\end{figure}

Finally with the n+${}^{12}$C system, in Fig.~\ref{Fig6} we show the even-parity
components of the cross sections but now when a small finite deformation
$\beta_2 = -0.2$ has been used. In this case only cross sections in the energy 
range 2.4 to 5.5 MeV are displayed which suffice to show three groupings of 
interest. The resonances seen in the total elastic cross section for this
(small) deformation in sequence as energy increases have 
spin-parities, $\frac{1}{2}^-$,
$\frac{5}{2}^+$, $\frac{3}{2}^+$, $\frac{7}{2}^+$, $\frac{5}{2}^+$,
$\frac{3}{2}^+$, $\frac{9}{2}^+$, and $\frac{1}{2}^+$.
The total cross section is depicted by the solid curve.  The notation for the 
separate component even-parity cross sections are as given for Fig.~\ref{Fig2}.
By comparison with results shown previously, we note that:
\begin{enumerate}
\item
The resonances can be assembled into 3 groups: an odd-parity singlet, an 
even-parity doublet and an even-parity quintuplet.
\item
The $\frac{1}{2}^+$ resonance (n~=~4), though registered through the RI procedure 
but which was not evident as a peak in the cross section in Fig.~\ref{Fig1}, 
now is seen as a little bump in the cross section above 5 MeV. 
Of the even-parity components, the $\frac{1}{2}^+$ one which dominates the
lowest energy values, clearly identifies that highest energy resonance in this 
figure as having that spin-parity.
\item
The widths are smaller than those found in Fig.~\ref{Fig1} and as was also very 
evident in the 3-D diagram (Fig.~\ref{Fig4}). They all tend to zero as 
$\left| \beta_2 \right|$ is 
gradually taken to zero, so confirming the nature of all of the compound 
resonances analyzed. To illustrate that variation in the widths $\Gamma$, a
select set of those for some of the larger resonances are plotted versus the 
deformation $\beta_2$ in Fig.~\ref{Fig7}.  As foreseen by theory~\cite{Pi95}, 
the widths tend to zero as $\left| \beta_2 \right|^2$.  Note that in one case (n~=~7), 
$\Gamma_{\text{th}} (J^\pi = \frac{3}{2}^+)$ 
initially increases as $\left| \beta_2 \right|$ decreases.  This is due to the coupling 
being quite strong so that the physical regime is far from the unperturbed limit.
\item
We have verified that with the small value for deformation, the n~=~8 $\frac{5}{2}^-$ 
state
is a resonance, a bound state in the continuum. For the physical value of 
$\beta_2 = -0.52$, it is an actual sub-threshold bound state.
\end{enumerate}

\section{The $^{13}$N system}
\label{Sec3}

In analyzing the $^{13}$N system we have assumed charge symmetry, and so use 
the parameter values given in Table~\ref{Tab1} for the n+${}^{12}$C system to 
evaluate p+${}^{12}$C scattering and to assess properties of the sub-threshold 
compound nucleus ${}^{13}$N. Only a Coulomb interaction has been added.
For simplicity we have chosen the Coulomb potential to be that of a uniformly
charged sphere so that we have only one new parameter, the Coulomb radius $R_c$.
The best agreement with the experimental data has been obtained with the value 
$R_c = 2.4$ fm.

The results, centroid energies and widths, are listed in Table~\ref{Tab7} and 
compared therein with experimental data~\cite{Aj91}. 
\begin{table}
\caption{\label{Tab7} Comparison between experiment and theory, p+${}^{12}$C.}
\begin{ruledtabular}
\begin{tabular}{cccccc}
n & $J^ \pi$ & $E_{\text{exp}}$ & $\frac{1}{2}\Gamma_{\text{exp}}$(keV) 
& $E_{\text{th}}$ & $\frac{1}{2}\Gamma_{\text{th}}$(keV)
\\
\hline
1 & $\frac{1}{2}^-$ & $-$1.9435 & --- & $-$1.9104 & ---  \\
2 & $\frac{1}{2}^-$ & 6.9745 & 115 & 5.6391 & 17.8  \\
3 & $\frac{1}{2}^+$ & 0.4214 & 15.8 & $-$0.0158 & ---  \\
4 & $\frac{1}{2}^+$ & 8.3065 & 140 & 6.9911 & 995  \\
5 & $\frac{3}{2}^-$ & 1.5675 & 31 & 1.5793 & 11.1  \\
6 & $\frac{3}{2}^+$ & 4.9425 & 57.5 & 4.7280 & 44.5  \\
7 & $\frac{3}{2}^+$ & 5.9565 & 750 & 5.8942 & 653  \\
8 & $\frac{5}{2}^-$ & 5.4325 & 37.5 & 2.9281 & 3.38 $\times 10^{-3}$ \\
9 & $\frac{5}{2}^+$ & 1.6035 & 23.5 & 0.6379 & 0.899  \\
10 & $\frac{5}{2}^+$ & 4.4205 & 5.5 & 4.1794 & 8.86  \\
11 & $\frac{5}{2}^+$ & --- & --- & 6.5281 & 726  \\
12 & $\frac{7}{2}^+$ & 5.2115 & 4.5 & 5.1234 & 1.50  \\
13 & $\frac{9}{2}^+$ & 7.0565 & 140 & 6.8341 & 153  \\
14 & $\frac{5}{2}^+$ & 9.5865 & 215 & 9.7895 & 910 
\end{tabular}
\end{ruledtabular}
\end{table} 
The comparison is quite good with the mean square error $\mu$, as specified by 
Eq.~(\ref{msqE}), relative to 13 states of Table~\ref{Tab7} being 0.2563~MeV. 
The entry in row 11 has been excluded from the sum since there is no 
experimental counterpart to that state. The average of the differences between 
pairs of states in $^{13}$N and $^{13}$C with reference to the experimental 
and theoretical spectra respectively, are the following:
\begin{equation}
\left< E_{\text{exp}}({}^{13}{\rm N}) - E_{\text{exp}}({}^{13}{\rm C}) \right>
= 3.01 \ {\rm MeV} \quad ; \quad
\left< E_{\text{th}}({}^{13}{\rm N}) - E_{\text{th}}({}^{13}{\rm C}) \right>
= 2.54 \ {\rm MeV}\ .
\end{equation}
Thus, through coupling with the $2^+$ level, the bound
state underlying the n~=~1 entry generates the compound odd-parity doublet 
n~=~5 and 8 in the tabulation, and that underlying the  resonance for n~=~3  
generates the quasi-compound even-parity doublet n~=~6 and 10. Likewise, the 
resonance n~=~9 links to the quasi-compound even-parity quintuplet formed
by the n~=~4, 7, 11, 12, and 13 entries in Table~\ref{Tab7}. Finally, the 
n~=~1 bound state when coupled with the second $0^+$ level gives rise to the 
single compound resonance for n~=~2.

For confirmation, in Table~\ref{Tab9} we give the results of calculations 
of the p+${}^{12}$C system in the unperturbed limit ($\beta_2 = 0$) with 
the spin-spin term set to zero.
\begin{table}
\caption{\label{Tab9} Sub-threshold bound states and resonances in
p+${}^{12}$C when $\beta_2 = 0$ and $V_{ss} = 0$.} 
\begin{ruledtabular}
\begin{tabular}{ccccc}
n & $J^ \pi$  & $E_{\text{th}}$ (MeV) & $\Delta E_n$ 
& $\frac{1}{2}\Gamma_{\text{th}}$ (keV) \\ 
\hline
1 & $\frac{1}{2}^-$ & $-$1.6785 & 0 & --- \\
2 & $\frac{1}{2}^-$ & 5.9757 & 7.6542 & 0 \\
3 & $\frac{1}{2}^+$ & 1.0217 & 0 & 468  \\
4 & $\frac{1}{2}^+$ &  6.6841 & 4.4389 & 115  \\
5 & $\frac{3}{2}^-$ & 2.7604 & 4.4389 & 0  \\
6 & $\frac{3}{2}^+$ &  5.4606 & 4.4389 & 467  \\
7 & $\frac{3}{2}^+$ &  6.6841 & 4.4389 & 115  \\
8 & $\frac{5}{2}^-$ & 2.7603 & 4.4388 & 0  \\
9 & $\frac{5}{2}^+$ & 2.2452 & 0 & 115  \\
10 & $\frac{5}{2}^+$ & 5.4606 & 4.4389 & 466  \\
11 & $\frac{5}{2}^+$ & 6.6841 &  4.4389 & 115  \\
12 & $\frac{7}{2}^+$ & 6.6841 & 4.4389 & 115  \\
13 & $\frac{9}{2}^+$ & 6.6841 & 4.4389 & 115 \\
14 & $\frac{5}{2}^+$ & 9.8995 &  7.6543 & 115
\end{tabular}
\end{ruledtabular}
\end{table}
The fourth column shows the differences which are consistent with the couplings
we suggest.  All the members of each multiplet tend to the same limit, and the 
difference between this limit and the relative {\it generator} single
particle state corresponds to the energy of the relevant
target excited state involved in the coupling. But there are some differences.
For example, the proton calculations predict four $\frac{5}{2}^+$ resonances 
in the energy range selected. That is one more than found from the neutron 
calculations in an equivalent energy range. Also, while the $\frac{5}{2}^+$
resonances n~=~9, 10 and 11
are the single particle state, a component of the even-parity doublet,  and
a component of the even-parity quintuplet as found in the neutron case, the 
n~=~14 resonance is new.  Our $\beta_2 \rightarrow 0$ analysis indicates that this 
(extra) state originates from the coupling to the excited $0^+_2$ state in 
${}^{12}$C.

An analysis of the widths listed in Table~\ref{Tab9} is interesting.
In the past~\cite{Pi95}, and derived from a purely mathematical example,
it was noted that the width of the compound resonance approaches zero 
as the coupling approaches zero (because its origin was a bound state), 
while the width of the quasi-compound ones tend to the natural width of the 
resonance from which such originated.  It is evident in Table~\ref{Tab9} that 
the odd-parity doublet (n~=~5 and 8) and the odd-parity resonance (n~=~2), 
are compound resonances as their widths vanish in the limit of zero deformation.
The 
even-parity quintuplet (n~=~4, 7, 11, 12, and 13) and the resonance (n~=~14)
have the width of the common resonance partner (n~=~9) and so are quasi-compound
resonances.  The width of the quasi-compound even-parity doublet (n~=~6 and 10)
is that of the partner resonance for which n~=~3.

Thus there is a one to one correspondence between the states
calculated by the model and those experimentally known, with the exception
of the $\frac{5}{2}^+$ resonance (n~=~14). But that resonance, while missing in 
n+$^{12}$C experimental data, has an observed  partner in p+${}^{12}$C system. 
Also, while it is clear from the comparisons to be made within
Table~\ref{Tab7} that the calculated energy centroids generally are in good 
agreement with data and that the widths concur in order of magnitude, there are
some exceptions.  In particular, the $\frac{5}{2}^-$ (n~=~8) state is not well 
reproduced. Also state n~=~3 is very close to threshold and, in spite of the 
small difference between theory and experiment, calculations make it be a 
sub-threshold bound state instead of a resonance.

Of course we have adopted a strict charge symmetry assumption and have used
a very simple form for the Coulomb interaction. It should be possible to
find improved ${}^{13}$N properties from the MCAS approach with but
slight changes of these potentials.

\subsection{Proton scattering from ${}^{12}$C; cross section and analyzing power}

Since Coulomb scattering amplitudes diverge at zero degree scattering,
measurements of proton scattering do not lead to total elastic scattering
cross sections.  Instead the usual procedure is to find cross sections at
fixed scattering angles and/or differential cross sections at fixed energies.
However, as the MCAS approach yields complete scattering (S-)matrices, 
such angular observables are readily predicted.

\begin{figure}
\scalebox{0.75}{\includegraphics*{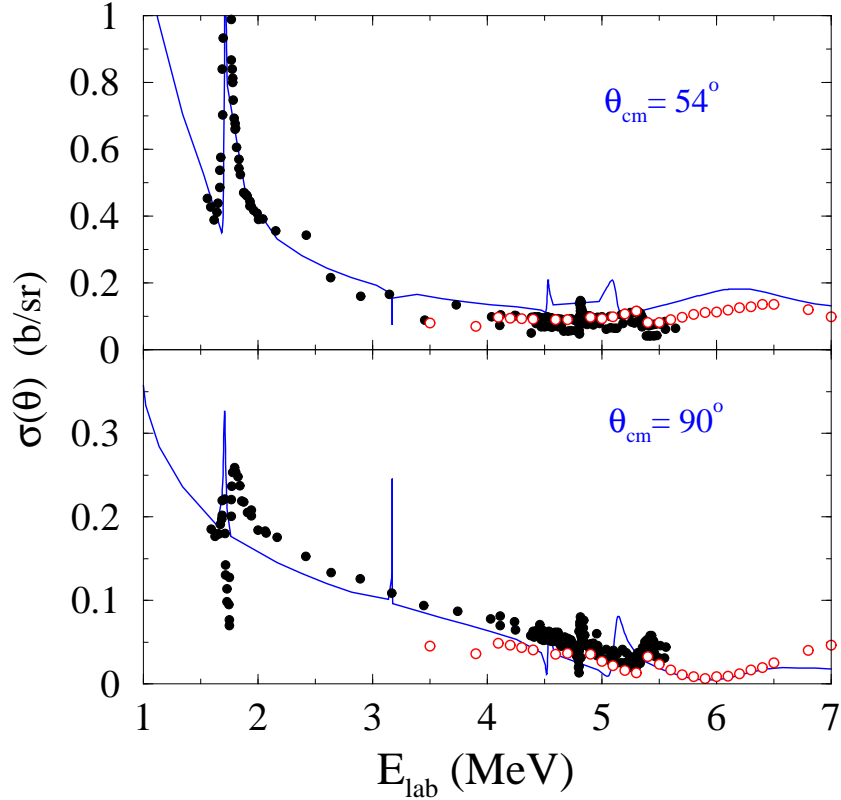}}
\caption{ \label{Fig8}(Color online)
Cross sections from proton elastic scattering from ${}^{12}$C for two
(cm) scattering angles, 54$^\circ$ (top panel), 90$^\circ$ (bottom panel).
Details are given in the text.}
\end{figure}

For energies to 7 MeV, proton elastic scattering from ${}^{12}$C at fixed 
(cm) scattering angles of 54$^\circ$ and at 90$^\circ$ are shown in the top 
and bottom panels of Fig.~\ref{Fig8} respectively. Therein our calculated
cross sections, shown by the continuous lines, are compared with 
available data.  With energies in the laboratory frame, the data measured at 
54$^\circ$ were taken from Ref.~\cite{Sy93} (open circles) while those
depicted by filled black circles come from Ref.~\cite{Re56}. 
At 54$^\circ$, the $\frac{3}{2}^-$ (n~=~5) resonance in the data is very well 
reproduced. The other resonances, $\frac{5}{2}^+$ (n~=~10), $\frac{3}{2}^+$ (n~=~6),
and the broad $\frac{3}{2}^+$ (n~=~7), also are well reproduced in shape and 
require but a small shift in energy. That need for a small energy shift is also
evident in their centroid values that are listed in Table~\ref{Tab7}. The sharp
peak shown in the theoretical curves near 3.2 MeV is the $\frac{5}{2}^-$ (n=8)
resonance. This state has been experimentally detected at higher energies,
These findings are confirmed 
by comparison of the calculated results with the data taken at 90$^\circ$ 
and displayed in the bottom panel.

\begin{figure}
\scalebox{0.75}{\includegraphics*{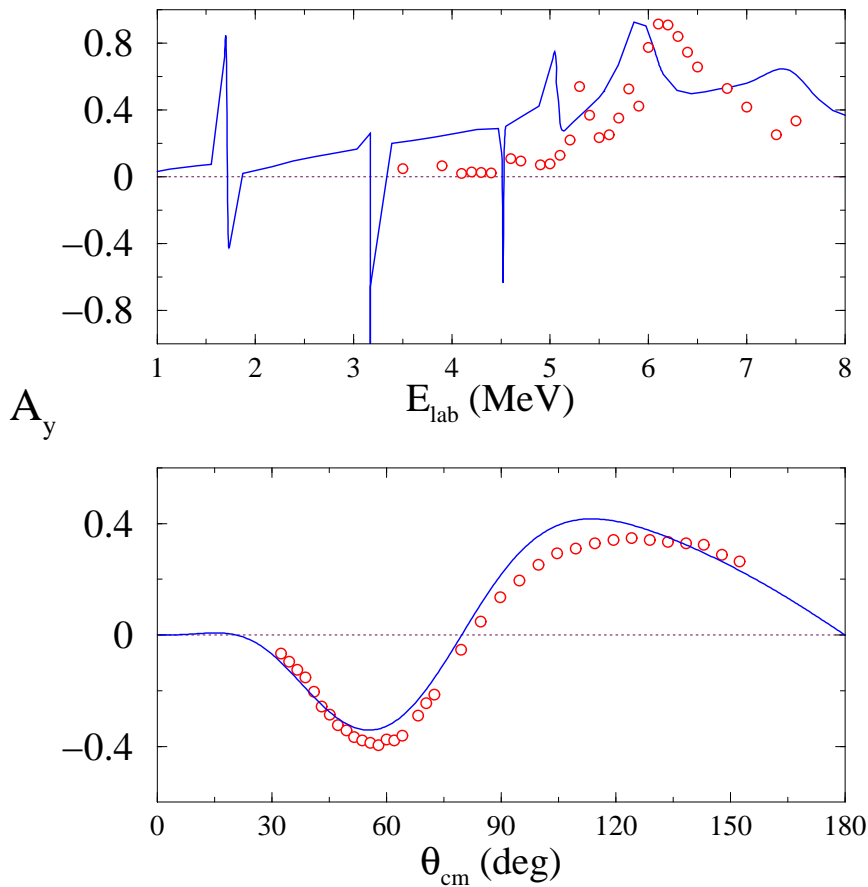}}
\caption{\label{Fig9}(Color online)
Experimental analyzing powers from proton scattering from ${}^{12}$C compared
with our theoretical calculations. Data and results taken for a range of
energies at a fixed scattering angle of 90$^\circ$ are shown in the top panel,
while angular variations at a fixed energy (of 3.5 MeV) are shown in the bottom 
panel.} 
\end{figure}

Usually analyzing powers $A_y$ from nucleon-nucleus scattering are more 
sensitive to details of structure than are the differential cross sections.
Two results for this observable are presented in Fig.~\ref{Fig9}. 
In the top panel of this figure, 
analyzing powers for proton elastic scattering at 90$^\circ$ (cm) and for
(lab) energies 1 to 8 MeV are depicted. In the bottom panel, angular variation
of the analyzing power for a fixed energy of 3.5 MeV is shown. 
The experimental data~\cite{Sy93} are depicted by open circles and our
calculated results are given by the solid curve.  The resonances seen 
in this figure are those that were discussed in relation to Fig.~\ref{Fig8}.
The bottom panel gives an angular variation of analyzing powers for an energy
(3.5 MeV) that is well removed from any strong resonance influence.
The calculated result then reflects what the model predicts as background
effects.  The agreement with data is good. Note that no additional 
parameter variation was done to achieve this agreement.

\begin{figure}
\scalebox{0.75}{\includegraphics*{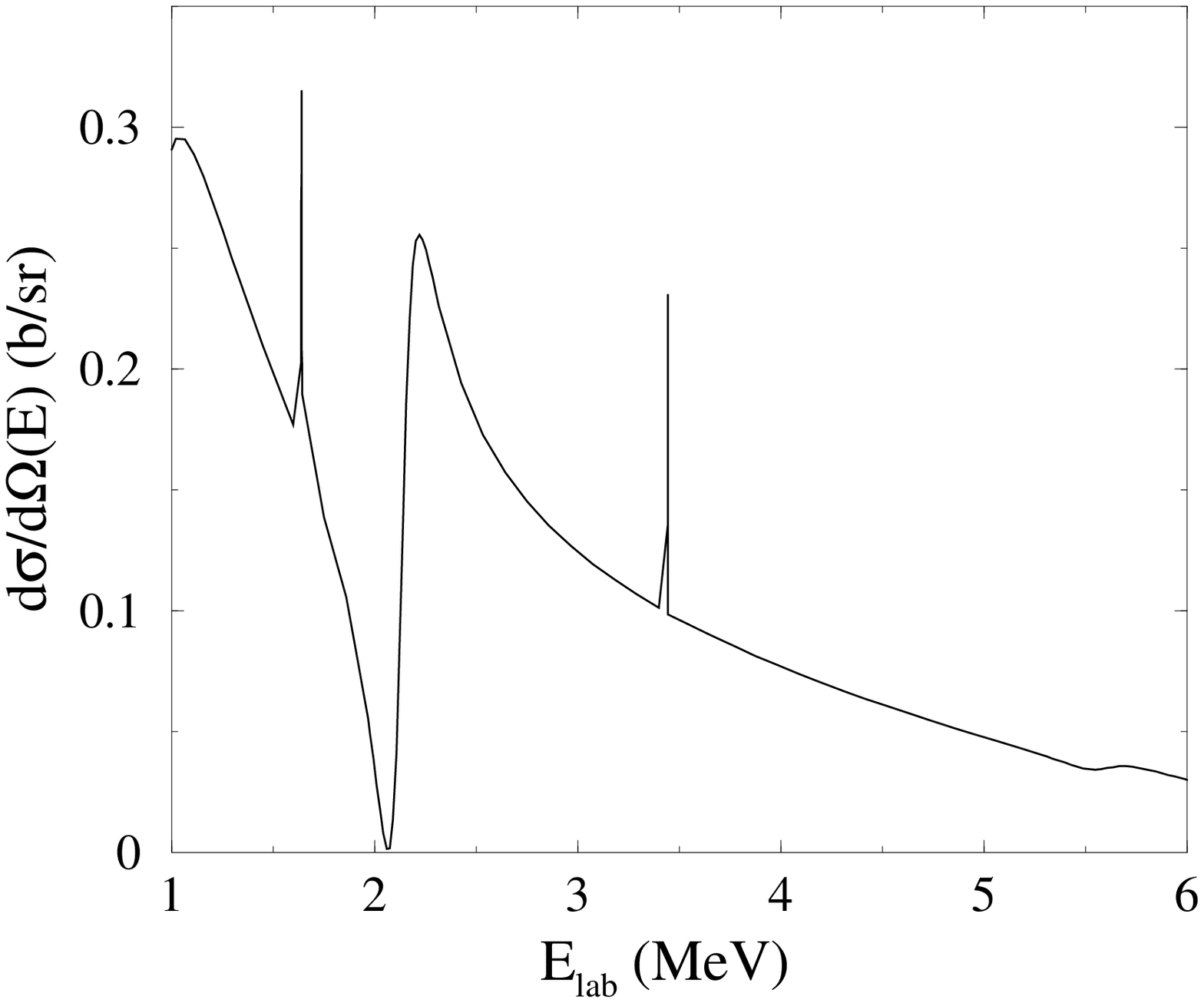}}
\caption{ \label{Fig10}
Energy variation of the theoretical differential cross section for proton 
scattering from ${}^{12}$C at a cm scattering angle of 90$^\circ$. The
calculation was made using $\beta_2 = -0.2$.} 
\end{figure}

To investigate deformation effects on differential cross sections, we 
calculated them again  using the values of parameters as listed in 
Table~\ref{Tab1} but with the deformation parameter $\beta_2 = -0.2$.
The results are shown in Fig.~\ref{Fig10} for a (cm) scattering angle of 
90$^\circ$ and (lab) energies ranging between 1 and 6 MeV.
This figure reveals how the single particle and compound resonances behave 
with decreased deformation. The $\frac{1}{2}^+$ (n~=~3) and  $\frac{5}{2}^+$ (n~=~9)
single particle resonances (the first of which was not seen in the preceding 
figure because it lies too low in energy) maintain their natural width, while 
the $\frac{3}{2}^-$ (n~=~5) and  $\frac{5}{2}^-$ (n~=~8) compound resonances 
sharpen. With $\beta_2 = -0.2$, already they are very narrow. The small bump 
between 5 and 6 MeV corresponds to the $\frac{5}{2}^+$ (n~=~10) resonance which 
maintains its natural width but, being  quasi-compound, is strongly damped as 
the unperturbed limit is approached.  Thus the behavior of the three kinds of 
resonances (single particle, compound and quasi-compound), as the coupling 
constant is decreased is just that foreseen previously~\cite{Pi95}. Namely,
as $\beta$ approaches zero, single particle resonances conserve their shapes,
compound resonances reduce their widths gradually keeping constant heights, while
quasi-compound resonances reduce their heights gradually keeping constant widths.

\section{Conclusions}
\label{Sec4}

We have analyzed the low energy spectra of $^{13}$C and $^{13}$N 
and low energy data from the elastic scattering of neutrons and of protons 
from ${}^{12}$C.  Our method of analysis was that of a multi-channel 
algebraic scattering theory with which both the (positive energy)
scattering of nucleons from ${}^{12}$C and (by using
negative energies) the sub-threshold bound states of the compound
nuclei could be predicted.  
The method, based upon sturmian expansions of a matrix of 
interaction potentials, ensures that all sub-threshold as well as
resonant states within the chosen range of energies are found. Also the
approach
can be adapted so that the Pauli principle is not violated even when a
collective model is used to define those interaction potentials.
In the cases studied, just such a rotational model was used for that purpose
and the $0^+$ (ground), $2^+$ (4.438 MeV), and the $0^+_2$ (7.96 MeV) states
in ${}^{12}$C taken as active.

\begin{figure}
\scalebox{0.65}{\includegraphics*{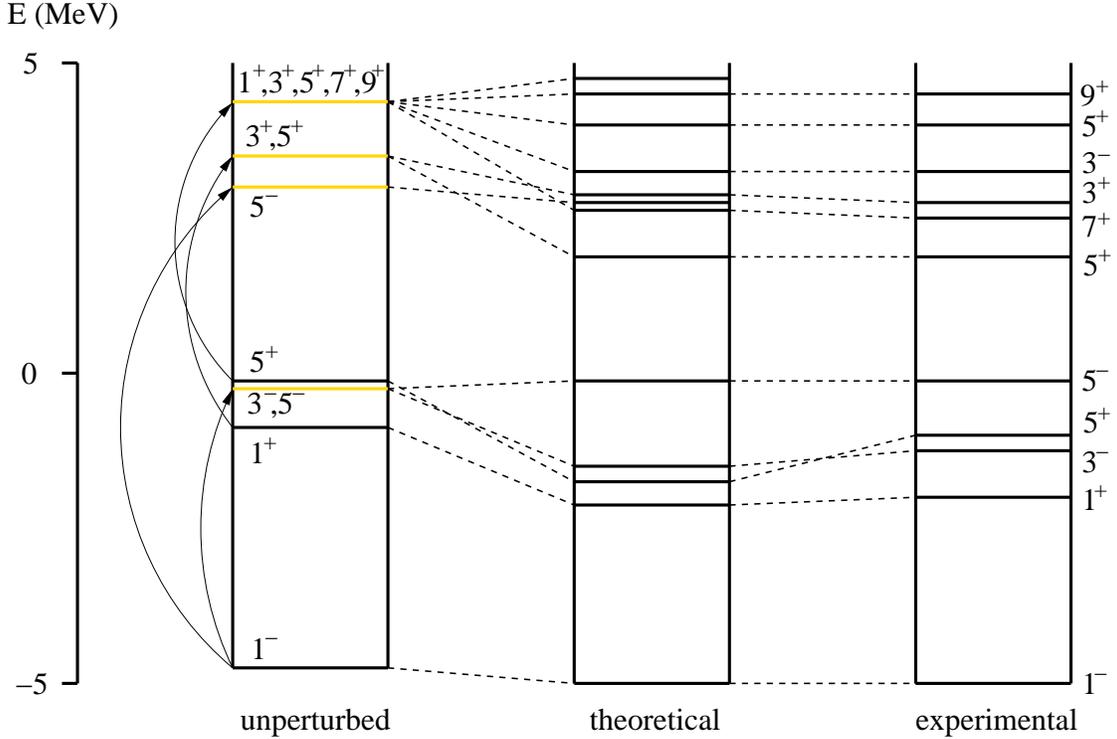}}
\caption{\label{Fig11}(Color online)
The genesis of the theoretical spectrum for ${}^{13}$C schematically 
represented and compared with the experimental spectrum.}
\end{figure}

\begin{figure}
\scalebox{0.6}{\includegraphics*{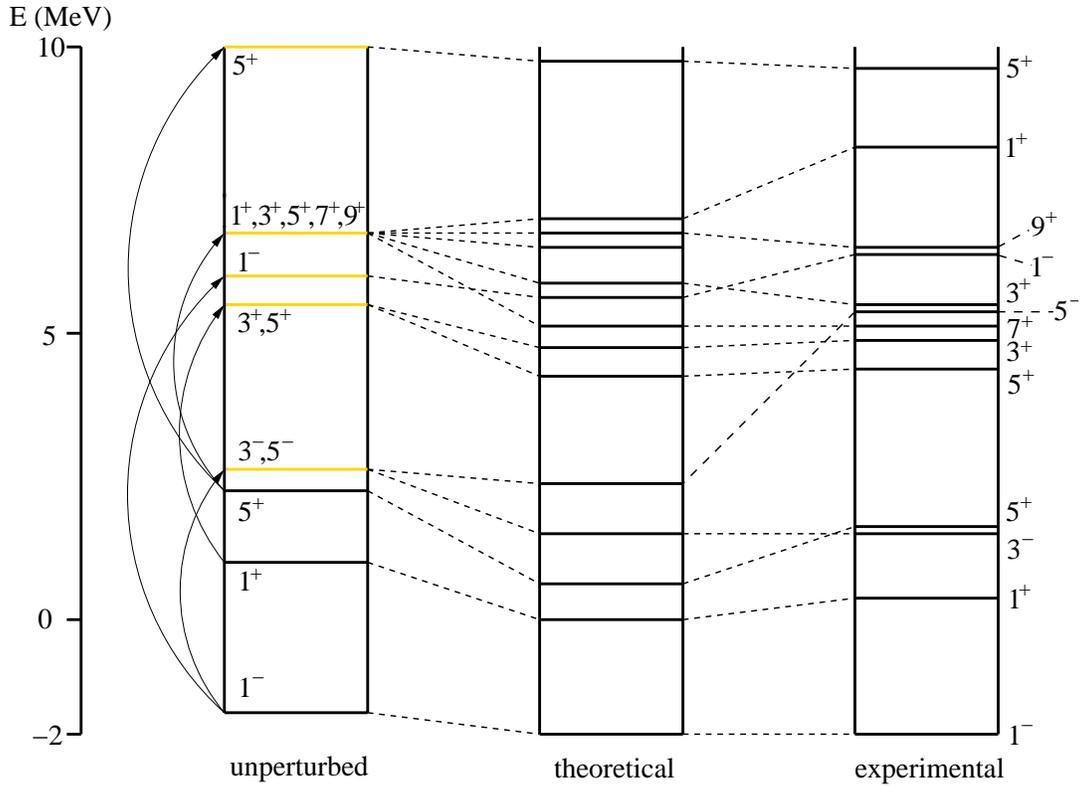}}
\caption{ \label{Fig12}(Color online)
The genesis of the theoretical spectrum for ${}^{13}$N schematically 
represented and compared with the experimental spectrum.}
\end{figure}

The results of our analysis are summarized in two figures, Fig.~\ref{Fig11}
and~\ref{Fig12}, for the compound nuclei ${}^{13}$C and ${}^{13}$N respectively.
Each displays three ladder diagrams; the first two are the results of our calculations
(unperturbed resulting when $\beta_2 = 0$) and they are compared with the 
experimental values (far right).
Considering Fig.~\ref{Fig11} for  ${}^{13}$C first, we note that it supports 
three (dominantly single particle) bound states, as is clear from the 
unperturbed 
($\beta_2 = 0$) spectrum depicted on the left of the figure. 
Those states are highlighted in that spectrum as dark lines. In addition the
coupling of the incoming neutron with the excited levels in the target 
${}^{12}$C then gives rise to meta-stable states whose presumed unperturbed 
($\beta_2 = 0$) configurations are represented by lighter lines.  They are
connected with what we deduced as a partner by the curved lines. The energy
gaps between these are $\epsilon_2$ and $\epsilon_3$ as relevant for the two 
possible cases. These unperturbed resonances are infinitely narrow 
($\Gamma_{\text{th}} \rightarrow 0$) and fully degenerate when one neglects the 
small effect of the 
spin-spin interaction in this representation.  The finite deformation 
($\beta_2 = -0.52$) splits these components to yield the predicted
resonances as shown in the middle of the set and which compares very favorably
with the experimental spectrum given in the box to the right.
Similar  conclusions can be drawn from the spectra for ${}^{13}$N that
are shown in Fig.~\ref{Fig12}.  
Because of the Coulomb energy shift, there
is only one sub-threshold bound state, and the MCAS approach assuming
charge symmetry gives just that and with the correct spin-parity.
As with the ${}^{13}$C system, accounting for the Pauli 
principle is crucial. Otherwise numerous spurious levels result.
An interesting difference however is that the Coulomb effect transforms 
the $\frac{1}{2}^+$ and $\frac{5}{2}^+$ bound states to be single particle 
resonances, and their products due to coupling with excited core states
become  quasi-compound resonances.
The different behavior of compound and quasi-compound resonances in the limit
$\beta_2 \rightarrow 0$ confirms statements formulated in a previous paper~\cite{Pi95}.

We conclude noting that, with a unique set of potential parameters the MCAS 
approach reproduces data.The results facilitate interpretation of the phenomenology 
of both the $^{13}$C and $^{13}$N systems in the considered energy range in 
both a satisfactory and a fairly exhaustive way. 


\begin{acknowledgments}
This research was supported by a grant from the Australian Research Council,
by a merit award with the Australian Partners for Advanced Computing, by the
Italian MURST-PRIN Project ``Fisica Teorica del Nucleo e dei Sistemi a Pi\`u
Corpi'', and by the Natural Sciences and Engineering Research Council (NSERC),
Canada. JPS, KA, and DvdK acknowledge the hospitality and support of the INFN,
Padova, and of the Dipartimento di Fisica, Universit\`a di Padova between 1999
and 2004. LC and JPS also would like to thank the School of Physics, University 
of Melbourne, as do LC, GP, and KA for the hospitality and support of the 
Department of Physics and Astronomy, University of Manitoba.
\end{acknowledgments}
 

\bibliography{compound-all}

\begin{thebibliography}{11}
\expandafter\ifx\csname natexlab\endcsname\relax\def\natexlab#1{#1}\fi
\expandafter\ifx\csname bibnamefont\endcsname\relax
  \def\bibnamefont#1{#1}\fi
\expandafter\ifx\csname bibfnamefont\endcsname\relax
  \def\bibfnamefont#1{#1}\fi
\expandafter\ifx\csname citenamefont\endcsname\relax
  \def\citenamefont#1{#1}\fi
\expandafter\ifx\csname url\endcsname\relax
  \def\url#1{\texttt{#1}}\fi
\expandafter\ifx\csname urlprefix\endcsname\relax\def\urlprefix{URL }\fi
\providecommand{\bibinfo}[2]{#2}
\providecommand{\eprint}[2][]{\url{#2}}

\bibitem[{\citenamefont{Amos et~al.}(2003)\citenamefont{Amos, Canton, Pisent,
  Svenne, and van~der Knijff}}]{Am03}
\bibinfo{author}{\bibfnamefont{K.}~\bibnamefont{Amos}},
  \bibinfo{author}{\bibfnamefont{L.}~\bibnamefont{Canton}},
  \bibinfo{author}{\bibfnamefont{G.}~\bibnamefont{Pisent}},
  \bibinfo{author}{\bibfnamefont{J.~P.} \bibnamefont{Svenne}},
  \bibnamefont{and} \bibinfo{author}{\bibfnamefont{D.}~\bibnamefont{van~der
  Knijff}}, \bibinfo{journal}{Nucl.\ Phys.} \textbf{\bibinfo{volume}{A728}},
  \bibinfo{pages}{65} (\bibinfo{year}{2003}).

\bibitem[{\citenamefont{Canton et~al.}(1988)\citenamefont{Canton, Cattapan, and
  Pisent}}]{Ca88}
\bibinfo{author}{\bibfnamefont{L.}~\bibnamefont{Canton}},
  \bibinfo{author}{\bibfnamefont{G.}~\bibnamefont{Cattapan}}, \bibnamefont{and}
  \bibinfo{author}{\bibfnamefont{G.}~\bibnamefont{Pisent}},
  \bibinfo{journal}{Nucl.\ Phys.} \textbf{\bibinfo{volume}{A487}},
  \bibinfo{pages}{33} (\bibinfo{year}{1988}).

\bibitem[{\citenamefont{Canton et~al.}(2004)\citenamefont{Canton, Pisent,
  Svenne, van~der Knijff, Amos, and Karataglidis}}]{Ca04}
\bibinfo{author}{\bibfnamefont{L.}~\bibnamefont{Canton}},
  \bibinfo{author}{\bibfnamefont{G.}~\bibnamefont{Pisent}},
  \bibinfo{author}{\bibfnamefont{J.~P.} \bibnamefont{Svenne}},
  \bibinfo{author}{\bibfnamefont{D.}~\bibnamefont{van~der Knijff}},
  \bibinfo{author}{\bibfnamefont{K.}~\bibnamefont{Amos}}, \bibnamefont{and}
  \bibinfo{author}{\bibfnamefont{S.}~\bibnamefont{Karataglidis}}
  (\bibinfo{year}{2004}), \bibinfo{note}{nucl-th/0409050, submitted for
  publication to Phys. Rev. Lett.}

\bibitem[{\citenamefont{Kukulin and Pomerantsev}(1978)}]{Ku78}
\bibinfo{author}{\bibfnamefont{V.}~\bibnamefont{Kukulin}} \bibnamefont{and}
  \bibinfo{author}{\bibfnamefont{V.}~\bibnamefont{Pomerantsev}},
  \bibinfo{journal}{Ann. of Phys.} \textbf{\bibinfo{volume}{111}},
  \bibinfo{pages}{330} (\bibinfo{year}{1978}).

\bibitem[{\citenamefont{Saito}(1969)}]{Sa69}
\bibinfo{author}{\bibfnamefont{S.}~\bibnamefont{Saito}},
  \bibinfo{journal}{Prog. Theor. Phys.} \textbf{\bibinfo{volume}{41}},
  \bibinfo{pages}{705} (\bibinfo{year}{1969}).

\bibitem[{\citenamefont{Pearlman}(1993)}]{Pe93}
\bibinfo{author}{\bibfnamefont{S.}~\bibnamefont{Pearlman}},
  \bibinfo{title}{ENDF/HE-VI mat-625} (\bibinfo{year}{1993}),
  \bibinfo{note}{BNL-48035}.

\bibitem[{\citenamefont{IAEA-NDS}(2005)}]{IA05}
\bibinfo{author}{\bibnamefont{IAEA-NDS}}, \emph{\bibinfo{title}{Computer index
  of neutron data \mbox{\rm (CINDA)}}} (\bibinfo{year}{2005}),
  \bibinfo{note}{International Atomic Energy Agency - Nuclear Data Services
  database version of January 4}.

\bibitem[{\citenamefont{Ajzenberg-Selove}(1991)}]{Aj91}
\bibinfo{author}{\bibfnamefont{F.}~\bibnamefont{Ajzenberg-Selove}},
  \bibinfo{journal}{Nucl.\ Phys.} \textbf{\bibinfo{volume}{A523}},
  \bibinfo{pages}{1} (\bibinfo{year}{1991}).

\bibitem[{\citenamefont{Pisent and Svenne}(1995)}]{Pi95}
\bibinfo{author}{\bibfnamefont{G.}~\bibnamefont{Pisent}} \bibnamefont{and}
  \bibinfo{author}{\bibfnamefont{J.~P.} \bibnamefont{Svenne}},
  \bibinfo{journal}{Phys.\ Rev.\ C} \textbf{\bibinfo{volume}{51}},
  \bibinfo{pages}{3211} (\bibinfo{year}{1995}), \bibinfo{note}{and references
  cited therein}.

\bibitem[{\citenamefont{Sydow et~al.}(1993)}]{Sy93}
\bibinfo{author}{\bibfnamefont{L.}~\bibnamefont{Sydow}} \bibnamefont{et~al.},
  \bibinfo{journal}{N.I.M.} \textbf{\bibinfo{volume}{327}},
  \bibinfo{pages}{441} (\bibinfo{year}{1993}).

\bibitem[{\citenamefont{Reich et~al.}(1956)\citenamefont{Reich, Phillips, and
  Russell}}]{Re56}
\bibinfo{author}{\bibfnamefont{C.~W.} \bibnamefont{Reich}},
  \bibinfo{author}{\bibfnamefont{G.~C.} \bibnamefont{Phillips}},
  \bibnamefont{and} \bibinfo{author}{\bibfnamefont{J.~L.}
  \bibnamefont{Russell}}, \bibinfo{journal}{Phys. Rev.}
  \textbf{\bibinfo{volume}{104}}, \bibinfo{pages}{143} (\bibinfo{year}{1956}).

\end{thebibliography}

\end{document}